\documentclass{PoS}
\usepackage{array}

\usepackage{slashed}

\newcommand{\be}{\begin{equation}}
\newcommand{\ee}{\end{equation}}
\newcommand{\bea}{\begin{eqnarray}}
\newcommand{\eea}{\end{eqnarray}}

\title{Spin Physics and Transverse Structure}

\ShortTitle{Spin Physics and Transverse Structure}

\author{\speaker{P.J. MULDERS}\\
        Nikhef Theory Group and Department of Physics and Astronomy, VU University Amsterdam\\
        De Boelelaan 1081, NL-1081 HV Amsterdam, the Netherlands\\
        E-mail: \email{mulders@few.vu.nl}}

\abstract{Spin is a welcome complication in the study of partonic structure that has led to new insights, even if theoretically and experimentally not all dust has settled, in particular on quark flavor dependence and gluon spin. At the same time it opened new questions on angular momentum and effects of transverse structure. In this talk the focus is on the role of the transverse momenta of partons. Like for collinear parton distribution functions (PDFs), we are also in the case of transverse momentum dependent (TMD) PDFs, talking about forward matrix elements. TMD PDFs (or in short TMDs) extend collinear PDFs with only spin-spin correlations to PDFs that include spin-momentum correlations, including also time-reversal-odd (T-odd) correlations, relevant for the description of single spin asymmetries. In this way TMDs open up new ways of studying the spin structure. Their operator structure within QCD, however, is more complex leading to various ways of breaking of universality.
}

\FullConference{XXIII International Workshop on Deep-Inelastic Scattering and Related Subjects\\
                 27 April - 1 May, 2015\\
                 Dallas, USA}

\begin{document}

\section{Introduction}
Parton densities are a natural way of dealing with the notion of cross sections as an incoherent sum of scattering off the partons, quarks and gluons, in a hadron. They are not needed in electroweak cross sections for leptons where the targets and produced particles are the degrees of freedom of the lagrangian describing its interactions. Then one just needs the wave functions including spinors or polarization vectors to account for external states, plane waves including their possible polarization. At higher orders the calculations involves of course the whole regularization and renormalization procedures, but these can be handled in a renormalizable theory. 

The wave functions and spinors or polarizations of partons actually show up in the single particle matrix elements of the corresponding fields. For a proper treatment of parton densities, one needs to consider the matrix elements of the fields between hadronic states, e.g.\ a single hadronic target in inclusive deep inelastic scattering (DIS), in the cross section leading to forward matrix elements of combinations of field operators. For local operator combinations this one can still employ the full field theoretical machinery. Such matrix elements, however, are merely moments of parton densities. Helped by the kinematics in a high-energy process one can make a twist expansion by identifying the leading local operator combinations, which then can through a Mellin transform be identified with collinear parton distribution functions (collinear PDF's). The renormalization differs for each of the local (composite) operators involving for each of them anomalous dimensions, leading to multiplicative renormalization factors for the moments and a convolution of PDFs with splitting functions to account for scale dependence of the matrix elements and the parton densities.

This all works amazingly well for QCD, where the collinearity shows up in data, that exhibit jet structure linked to 'target remnants' and 'scattered partons'. There is a natural order of scales. Starting with the high energy scale of the process, say $\sqrt{s}$ and a fraction of it linked to the still high energy scale of the partonic process, say $Q$ to the transverse momenta of the order of a GeV or less. Intermediate are larger values of the transverse momenta but viewed as part of the partonic probabilities these fall like $\alpha_s(\vert p_T^2\vert)/\vert p_T^2\vert$. 

Of course high energy kinematics is an essential ingredient in this. The hadronic mass becomes irrelevant and a given hadron is struck at an instant (light-front time). If $P_1\cdot P_2 \rightarrow \infty$ one has $s\rightarrow \infty$. Any momentum $p = x_1\,P_1 + x_2\,P_2 + p_T$ with $p_T{\cdot}P_1=p_T{\cdot}P_2 = 0$ and hadronic size $p^2 \sim p_T^2 \sim M^2$ naturally must have $x_1\,x_2 \sim M^2/s$ splitting into three types, $x_1$ finite and $x_2 \sim M^2/s\rightarrow 0$ ($p$ is momentum of a parton in 1), $x_1 \sim M^2/s\rightarrow 0$ and $x_2$ finite ($p$ is momentum of parton in 2), or $x_1 \sim x_2 \sim M/\sqrt{s} \rightarrow 0$ ({\em soft}). For a parton in hadron 1, one can then has a Sudakov decomposition
\[
p = x_1\,P_1 + p_T + \frac{1}{2}(p^2 - p_T^2)\,n,
\]
where $n = P_2/P_1{\cdot}P_2$. The approximate light-like vector $n$ satisfies $P_1{\cdot}n = 1$ and $x_1 = p{\cdot}n$. If another hadronic vector $K$ for which $P_1{\cdot}K \sim s$ is used to fix $n$, the change in $x_1$ is of order $M^2/s$. The intrinsic transverse momentum $p_T$ will depend on the choice of $n$. Considering a process like the Drell-Yan process, two partons of different hadrons combine into a hard final state momentum with $q^2 = Q^2$ (e.g.\ a lepton pair), the measurable quantity
\[
q = \frac{q{\cdot}P_2}{P_1{\cdot}P_2}\,P_1 + \frac{q{\cdot}P_1}{P_1{\cdot}P_2}\,P_2 + q_T,
\]
includes the experimentally accessible non-collinearity $q_T$. The rest identifies the partonic momentum fractions (up to mass corrections) with scaling variables,
\[
x_1 = \frac{q{\cdot}P_2}{P_1{\cdot}P_2} \approx \frac{Q^2}{2P_1{\cdot}q} 
\quad \mbox{and} \quad
x_2 = \frac{q{\cdot}P_1}{P_1{\cdot}P_2} \approx \frac{Q^2}{2P_2{\cdot}q} .
\]
The basic subprocess is parton($p_1$) + parton ($p_2$) $\rightarrow \gamma^\ast(q)$. The actual sensitivity to the transverse momentum shows up in the transverse momentum of the lepton pair. If $p_1 = x_1P_1 + p_{1T}$ and $p_2 = x_2P_2 + p_{2T}$ one has $q_T = p_{1T} + p_{2T}$ at leading order linking the partonic momenta to the observed non-collinearity $q_T$. The cross section in terms of the transverse momenta is a convolution in momentum space or a product in impact parameter space, for which we hope to write at leading order (in an expansion in $M^2/s$)
\begin{eqnarray}
d\sigma &\propto & \int d^2p_{1T}\,d^2p_{2T}\,\delta^2(p_{1T}+p_{2T}-q_T)\ldots 
\Phi(x_1,p_{1T};\ldots)\,\Phi(x_2,p_{2T};\ldots)\ldots
\nonumber\\
&\propto & \int d^2b_T \ \exp(i\,q_T{\cdot}b_T)\ldots 
\Phi(x_1;b_{T};\ldots)\,\Phi(x_2;b_{T};\ldots)\ldots .
\end{eqnarray}
Besides hard amplitudes and kinematic factors the dots (in particular those in the arguments of the correlators) also include regulators and corresponding scale dependence to handle large $p_T$ UV behavior as well as rapidity divergences~\cite{Collins:2011zzd}.
Effects of {\em intrinsic} transverse momenta of partons are best visible in (partially) polarised processes. In that case one has polarization vectors $S$ for hadrons (parametrizing the spin density matrix) or measurable polarization vectors depending on final state distributions of decay products, e.g.\ in $\rho \rightarrow \pi\pi$ or in $\Lambda \rightarrow \pi N$. These spin vectors can fix directions including in particular the transverse ones. The spin vector can be parametrized as $M\,S = S_L\,P + M\,S_T - M^2\,S_L\,n$, which obeys $S^2 = -S_L^2 + S_T^2$ and $P{\cdot}S = 0$. Therefore the polarization is not correlated with the collinear component of the momentum. Collinear quark densities, indeed, just come as spin-spin densities, unpolarized densities ($f_1^q(x)$ or $q(x)$) in an unpolarized proton, longitudinally polarized (chiral) densities ($g_1^q(x)$ or $\Delta q(x)$) in longitudinally polarized proton and transversely polarized densities ($h_1^q(x)$ or $\delta q(x)$) in a transversely polarized proton. Similarly gluon densities are just unpolarized gluon densities ($f_1^g(x)$ or $g(x)$) in an unpolarized nucleon or circularly polarized gluon densities ($g_1^g(x)$ or $\Delta g(x)$) in a longitudinally polarized nucleon.  

\section{TMD correlators and distribution functions}

The quark and gluon TMD correlators in terms of matrix elements of quark fields~\cite{Collins:1981uw,Collins:1981uk} including the Wilson lines $U$ needed for color gauge invariance of the TMD case are given by
\begin{eqnarray}
&&
\Phi_{ij}^{[U]}(x,p_T;n)
=\int \frac{d\,\xi{\cdot}P\,d^{2}\xi_T}{(2\pi)^{3}}
\,e^{ip\cdot \xi} \langle P{,}S\vert\overline{\psi}_{j}(0)
\,U_{[0,\xi]}\psi_{i}(\xi)\vert P{,}S\rangle\,\big|_{LF},
\label{e:operator}
\\&&
2x\,\Gamma^{[U,U^\prime]\,\mu\nu}(x,p_T;n) ={\int}\frac{d\,\xi{\cdot}P\,d^2\xi_T}{(2\pi)^3}\ e^{ip\cdot\xi}
\,\langle P{,}S\vert\,F^{n\mu}(0)\,U_{[0,\xi]}^{\phantom{\prime}}\,F^{n\nu}(\xi)\,U_{[\xi,0]}^\prime\,\vert P{,}S\rangle\big|_{LF}
\end{eqnarray}
(color summation or color tracing implicit), where the Sudakov decomposition for the momentum $p^{\mu}$ of the produced quark or gluon is used. The non-locality in the integration is limited to the lightfront, $\xi{\cdot}n = 0$, indicated with LF. The gauge links $U_{[0,\xi]}^{\phantom{\prime}}$ are path ordered exponentials needed to make the correlator gauge invariant~\cite{Belitsky:2002sm,Boer:2003cm}. Depending on the process under consideration different gauge links will appear~\cite{Bomhof:2006dp,Bomhof:2007xt}. For the quark correlator the gauge link bridges the non-locality, which in the case of TMDs involves also transverse separation. The simplest ones are the future- and past-pointing staple links $U_{[0,\xi]}^{[\pm]}$ (or just $[\pm]$) that just connect the points $0$ and $\xi$ via lightcone plus or minus infinity, $U_{[0,\xi]}^{[\pm]} = U_{[0,\pm\infty]}^{[n]} U_{[0_{\scriptscriptstyle T},\xi_{\scriptscriptstyle T}]}^{{\scriptscriptstyle T}}U_{[\pm \infty,\xi]}^{[n]}$. We use these as our basic building blocks. For gluons TMDs the most general structure involves two gauge links (triplet representation), denoted as $[U,U^\prime]$, connecting the positions $0$ and $\xi$ in different ways. The simplest combinations allowed for $[U,U^\prime]$ are $[+,+^\dagger]$, $[-,-^\dagger]$, $[+,-^\dagger]$ and $[-,+^\dagger]$. More complicated possibilities, e.g. with additional (traced) Wilson loops of the form $U^{[\square]}=U_{[0,\xi]}^{[+]}U_{[\xi,0]}^{[-]}$ = $U_{[0,\xi]}^{[+]}U_{[0,\xi]}^{[-]\dagger}$ or its conjugate are allowed as well. A list with all type of contributions can be found in Ref.~\cite{Buffing:2012sz,Buffing:2013kca}. If $U = U^\prime$ one can also use a single gauge link in the octet representation.

Since the above correlator cannot be calculated from first principles, an expansion in terms of TMD PDFs is used, which at the level of leading twist contributions is given by~\cite{Mulders:1995dh,Bacchetta:2006tn,Mulders:2000sh,Meissner:2007rx}
\begin{eqnarray}
\Phi^{[U]}(x,p_{T};n)&=&\bigg\{
f^{[U]}_{1}(x,p_T^2) + g^{[U]}_{1s}(x,p_T^2) + i\,h_1^{\perp [U]}(x,p_T^2)\frac{\slashed{p}_T}{M}
\nonumber \\&&
\mbox{}+h^{[U]}_{1}(x,p_T^2)\,\gamma_5\,\slashed{S}_{T}
+h_{1T}^{\perp [U]}(x,p_T^2)\,\frac{p_{T\alpha\beta}S_T^{\{\alpha}\gamma_T^{\beta\}}\gamma_5}{2M^2}
\bigg\}\frac{\slashed{P}}{2} .
\label{e:QuarkCorr}
\\
2x\,\Gamma^{\mu\nu [U]}(x{,}p_{\scriptscriptstyle T}) &=& 
-g_T^{\mu\nu}\,f_1^{g [U]}(x{,}p_{\scriptscriptstyle T}^2)
+g_T^{\mu\nu}\frac{\epsilon_T^{p_TS_T}}{M}\,f_{1T}^{\perp g[U]}(x{,}p_{\scriptscriptstyle T}^2)
\nonumber\\&&
\mbox{}+i\epsilon_T^{\mu\nu}\;g_{1s}^{g [U]}(x{,}p_{\scriptscriptstyle T})
+\bigg(\frac{p_T^\mu p_T^\nu}{M^2}\,{-}\,g_T^{\mu\nu}\frac{p_{\scriptscriptstyle T}^2}{2M^2}\bigg)\;h_1^{\perp g [U]}(x{,}p_{\scriptscriptstyle T}^2)
\nonumber\\ &&
\mbox{}-\frac{\epsilon_T^{p_T\{\mu}p_T^{\nu\}}}{2M^2}\;h_{1s}^{\perp g [U]}(x{,}p_{\scriptscriptstyle T})
-\frac{\epsilon_T^{p_T\{\mu}S_T^{\nu\}}{+}\epsilon_T^{S_T\{\mu}p_T^{\nu\}}}{4M}\;h_{1T}^{g[U]}(x{,}p_{\scriptscriptstyle T}^2).
\label{e:GluonCorr}
\end{eqnarray}
We have used that $S^\mu = S_{\scriptscriptstyle L}P^\mu + S^\mu_{\scriptscriptstyle T} + M^2\,S_{\scriptscriptstyle L}n^\mu$. For function like $g_{1s}^{[U]}$ and $h_{1s}^{\perp[U]}$ the shorthand notation
\begin{equation}
g_{1s}^{[U]}(x,p_{\scriptscriptstyle T})=S_{\scriptscriptstyle L} g_{1L}^{[U]}(x,p_{\scriptscriptstyle T}^2)-\frac{p_{\scriptscriptstyle T}\cdot S_{\scriptscriptstyle T}}{M}g_{1T}^{[U]}(x,p_{\scriptscriptstyle T}^2)
\end{equation}
is used.
The gauge link dependence in this parametrization is contained in the TMDs. Note that for quarks
$f_{1T}^{\perp}$ and $h_1^\perp$ are T-odd, while for gluons $f_{1T}^{\perp g}$, $h_{1T}^{g}$, $h_{1L}^{\perp g}$ and $h_{1T}^{\perp g}$ are T-odd. 

Even if any gauge link defines a gauge invariant correlator, the relevant gauge links to be used in a given process just follows from a correct resummation of all diagrams including the exchange of any number of $A^n$ gluons between the hadronic parts and the hard part, i.e.\ gluons with their polarization along the hadronic momentum. They nicely sum to the path-ordered exponential. For quark distributions in semi-inclusive deep inelastic scattering they resum into a future-pointing gauge link, in the Drell-Yan process they resum into a past-pointing gauge link, which is directly linked to the color flow in these processes.

\section{Operator analysis}

In the situation of collinear PDFs (integrated over transverse momenta), the non-locality is restricted to the lightcone, $\xi{\cdot}n = \xi_T = 0$ (LC) and the staple links reduce to straight-line Wilson lines. The correlators then involve the non-local operator combinations $\overline\psi(0)U^{[n]}_{[0,\xi]}\psi(\xi)\vert_{LC}$ or $F^{n\mu}(0)U^{[n]}_{[0,\xi]}F^{n\nu}(\xi)U^{[n]}_{[\xi,0]}\vert_{LC}$, expanded in terms of leading twist operators $\overline\psi(0)D^n\ldots D^n\psi(0)$ and ${\rm Tr}[F^{n\mu}D^n\ldots D^n F^{n\nu}(0)D^n\ldots D^n]$ operators. Although gauge links are part of the matrix elements, they do not cause any non-universality or process dependence. In order to find the local operators for TMDs we integrate over transverse momentum including explicit transverse momentum vectors. Each transverse momentum $p_T^\alpha$ becomes a derivative giving transverse indices. The simplest of these transverse moments are
\begin{eqnarray}
&&
\int d^2p_{\scriptscriptstyle T}\ \Phi^{[U]}(x,p_{\scriptscriptstyle T})=\widetilde\Phi(x),
\\&&
\int d^2p_{\scriptscriptstyle T}\ p_{\scriptscriptstyle T}^{\alpha}\,\Phi^{[U]}(x,p_{\scriptscriptstyle T})=\widetilde\Phi_{\partial}^{\alpha}(x)+C_{G,c}^{[U]}\,\widetilde\Phi_{G,c}^{\alpha}(x),
\\&&
\int d^2p_{\scriptscriptstyle T}\ p_{\scriptscriptstyle T}^{\alpha_1}p_{\scriptscriptstyle T}^{\alpha_2}\,\Phi^{[U]}(x,p_{\scriptscriptstyle T})=\widetilde\Phi_{\partial\partial}^{\alpha_1\alpha_2}(x)+C_{G,c}^{[U]}\,\widetilde\Phi_{\{\partial G\},c}^{\alpha_1\alpha_2}(x)+C_{GG,c}^{[U]}\,\widetilde\Phi_{GG,c}^{\alpha_1\alpha_2}(x),
\label{e:moments}
\end{eqnarray}
and similarly results for $\Gamma^{[U,U^\prime]}(x,p_T)$. These integrated results will require standard UV regularization and corresponding scale dependence, while one also may need to consider appropriate combinations, e.g.\ subtraction of traces, to get finite results. Furthermore it can often be more appropriate to work with Bessel moments~\cite{Boer:2011xd}. The correlators appearing in Eq.~\ref{e:moments} are of the form 
\bea
&&
\widetilde\Phi_{\hat O,ij}^{[U]}(x,p_{T})
=\int \frac{d\,\xi{\cdot}P\,d^{2}\xi_{T}}{(2\pi)^{3}}
\,e^{ip\cdot \xi} \langle P{,}S\vert\overline{\psi}_{j}(0)
\,U_{[0,\xi]}\hat O(\xi)\psi_{i}(\xi)\vert P{,}S\rangle\,\Big|_{LF},
\label{e:twistoperatorquark}
\eea
where the $\hat O(\xi)$ operators are rank two combinations of $i\partial_T(\xi) = iD_T^\alpha(\xi) - A_T^\alpha(\xi)$ and $G^\alpha(\xi)$, defined in a color gauge invariant way (thus including GLs),
\bea
&&A_{T}^{\alpha}(\xi)=\frac{1}{2}\int_{-\infty}^{\infty}
d\eta{\cdot}P\ \epsilon(\xi{\cdot}P-\eta{\cdot}P)
\,U_{[\xi,\eta]}^{[n]} F^{n\alpha}(\eta)U_{[\eta,\xi]}^{[n]}, 
\label{e:defA} 
\\
&&G^{\alpha}(\xi)=\frac{1}{2}\int_{-\infty}^{\infty}
d\eta{\cdot}P\ U_{[\xi,\eta]}^{[n]}F^{n\alpha}(\eta)
U_{[\eta,\xi]}^{[n]},
\label{e:defG}
\eea
with $\epsilon (\zeta)$ being the sign function. Note that $G^{\alpha}(\xi)$ = $G^{\alpha}(\xi_T)$ does not depend on $\xi{\cdot}P$, implying in momentum space $p\cdot n = p^+ = 0$, hence the name gluonic pole matrix elements~\cite{Efremov:1981sh,Efremov:1984ip,Qiu:1991pp,Qiu:1991wg,Qiu:1998ia,Kanazawa:2000hz}. In Eq.~\ref{e:moments} one encounters symmetrized products of these operators indicated with subscripts $\{\partial G\}$, etc. Moreover the color summation often introduces multiple possibilities, e.g.\ for a gluonic pole matrix element in combination with two gluon fields there are two possibilities, ${\rm Tr}(F[G,F])$ ($c=1$) and ${\rm Tr}(F\{G,F\})$ ($c = 2$) that have to be summed over. The final important ingredient in Eq.~\ref{e:moments} are the gluonic pole factors, calculable factors depending on the number of gluonic poles in the operator and the path of the gauge link $U$. Most well-known are the single gluonic pole factors $C_G^{[\pm]} = \pm 1$. 

The operators $\widetilde\Phi_{\hat O}(x)$ can be considered as zero momentum limits or integrations over of multi-parton correlators such as three-parton quark-gluon-quark correlators $\Phi_{D}(x,y)$ with an operator structure $\langle\overline\psi(0)D^\alpha(\eta)\psi(\xi)\rangle$, non-local along the lightcone, and similarly $\Phi_{F}(x,y)$ or gluon-gluon-gluon correlators like $\Gamma_{F}(x,y)$. These multi-parton distributions play a role in higher twist contributions to the cross sections and can be used to establish relations, they appear in sum rules, they exhibit symmetries between correlators involving partons and anti-partons, etc. Several contributions on these topics are presented at this conference. 

In many cases the multi-parton distributions are also driving the large $p_T$-behavior of the TMDs. For the collinear PDFs extended to the corresponding TMDs, this is the $\alpha_s/p_T^2$ behavior of $f_1(x,p_T^2)$, in essence the evolution equations. In this way multi-parton distributions are driving the large $p_T$ behavior of many of the TMDs~\cite{Bacchetta:2008xw,Echevarria:2015uaa}, in particular the T-odd ones. Note, however, that the large $p_T$ behavor of TMDs without a collinear counterpart may be driven by collinear functions, such as $f_1^g$ driving the linearly polarized gluon distribution $h_1^{\perp g}$~\cite{Catani:2010pd,Catani:2011kr} or the transversity $h_1$ driving the Pretzelocity function $h_{1T}^\perp$~\cite{Bacchetta:2008xw}.  

\section{Universal TMDs}

Taking transverse derivatives gives the coefficients in the expansion in transverse momenta, for which we like to use traceless irreducible tensors $p_T^{\alpha_1\alpha_2\ldots}$ of a fixed rank, which describe in essence the azimuthal dependence. We would like to use the moments to identify the coefficients in the azimuthal expansion
\begin{equation}
\Phi(x,p_T) = \widetilde\Phi(x,p_T^2) + \frac{p_{Ti}}{M}\,\widetilde\Phi^{i}(x,p_T^2) + \frac{p_{Tij}}{M^2}\,\widetilde\Phi^{ij}(x,p_T^2) + \ldots .
\end{equation}
If in the higher moments just operators of the type $\widetilde\Phi_{\partial\ldots\partial}^{\alpha_1\ldots\alpha_n}$ would appear, it would be easy to find the operator expressions for $f_1(x,p_T^2)$. It would correspond with the rank zero operator $\widetilde\Phi(x,p_T^2)$ including all $\partial{\cdot}\partial$ traces that are subtracted in the higher rank operators and account for the $p_T^2$ dependence. Such is actually the case for fragmentation functions where the gluonic pole matrix elements (after integration over transverse momenta) vanish. 

For the distribution correlators, however, this procedure does not lead to a unique correlator linked to a particular function because gluonic pole matrix elements do not vanish, in particular also terms including $G{\cdot}G$ traces. For instance even the unpolarized quark (or gluon) TMD distributions $f_1(x,p_T^2)$ remain gaugelink-dependent~\cite{Boer:2015kxa} because of this. Only the $\partial{\cdot}\partial$ traces are taken care of in the $p_T^2$-dependence of the function, but gluonic pole trace operators $\widetilde\Phi_{G{\cdot}G,c}(x,p_T^2)$ need to be included and require introduction of functions $\delta f_1^{[GG,c]}(x,p_T^2)$. These functions have to satisfy $\int d^2p_T\ p_T^2\,\delta f_1^{[GG,c]}(x,p_T^2) = 0$, so they cause gauge-link dependent modulations,
\begin{equation}
f_1^{[U]}(x,p_T^2) = f_1(x,p_T^2) + \sum_{c=1,2}C_{GG,c}^{[U]}\,\delta f_1^{[GG,c]}(x,p_T^2) + \ldots ,
\end{equation}
of which only the first term survives in the collinear, $p_T$-integrated, situation. It leads to process dependence in the $p_T^2$ dependence, e.g.\ in the $p_T$-width. In this case there are two possible color contractions in the summation over $c$.  There is no term with a single gluonic pole matrix element because of the time reversal nature of the gluonic poles being odd. The time reversal nature, however, makes T-odd functions gaugelink-dependent already from the start, such as for the Sivers function,
\begin{equation}
f_{1T}^{\perp[U]}(x,p_T^2) = C_{G}^{[U]}\,f_{1T}^{\perp[G]}(x,p_T^2) + \ldots .
\end{equation}
For the Pretzelocity distribution there are three different operator structures contributing to the gauge link dependence, 
\begin{equation}
h_{1T}^{\perp[U]}(x,p_T^2) = h_{1T}^{\perp [\partial\partial]}(x,p_T^2) 
+ \sum_{c=1,2}C_{GG,c}^{[U]}\,h_{1T}^{\perp[GG,c]}(x,p_T^2) + \ldots .
\end{equation}
Thus measurements of Pretzelocity effects are gaugelink-dependent, and hence process dependent, even if the observable is a T-even function. In any given process a particular combination of the universal functions on the righthandside appears. These universal functions are here labeled by a combination of $\partial$ and $G$ identifying the operator structure and if needed an index $c$ if multiple color configurations have to be considered. In all of the above expressions, one has to be aware of additional modulations that come from operators with even more (traced) gluonic pole terms. To study their possible importance lattice studies using different gaugelink structures would be useful~\cite{Musch:2011er,Engelhardt:2014wra}. 

Turning to the gluon distributions, one also encounters gaugelink-dependence already for the unpolarized TMD distributions,
\begin{eqnarray}
f_{1}^{g[U,U^\prime]}(x,p_T^2)&=&f_1^g(x,p_T^2) 
+ \sum_{c=1}^4 C_{GG,c}^{[U,U^\prime]}\,\delta f_{1}^{g[GG,c]}(x,p_{T}^2) + \ldots, 
\end{eqnarray}
including four different color configurations in the $p_T^2$ modulation. Like for the Pretzelocity, one has for a T-even situation such as that of linearly polarized gluons in an unpolarized hadron also multiple universal functions,
\begin{eqnarray}
h_1^{\perp g[U,U^\prime]}(x,p_T^2)&=&h_1^{\perp g [\partial\partial]}(x,p_T^2)
+\sum_{c=1}^{4}C_{GG,c}^{[U,U^\prime]}\,h_1^{\perp g [U,U^\prime]}(x,p_T^2) + \ldots .
\end{eqnarray}
The above examples illustrate our ongoing efforts~\cite{Buffing:2013kca} to establish a universal set of TMD functions. As a final note, I want to stress that in situations where two (or more) TMDs with nonzero rank are involved, one must account for possible additional color factors in the basic expressions that are in DY-like processes for instance different from the $1/N_c$ or $1/(N_c^2-1)$ factors for $q\overline q$ or $gg$ initiated processes~\cite{Buffing:2011mj,Buffing:2013dxa}.

\section{Conclusions}

TMDs encode many features that can be linked to the partonic structure of hadrons and that can in principle be accessed at leading order provided that one picks the right variable, usually involving azimuthal asymmetries in polarized processes. Such efforts are under investigation in the experimental programs at RHIC/Brookhaven, JLab, BELLE, COMPASS/CERN, JPARC, BESIII or BaBar (see contributions at this workshop). In parallel, theoretical developments are underway to understand the data and the way these have to be incorporated into our view of the partonic structure of hadrons. This is a nontrivial enterprise since there are many ideas, but also many technical hurdles to take. In my talk I have focussed for a large part of efforts in establishing a universal set, connected to specific operators, for TMDs that one can then can try to work with~\cite{Hautmann:2014kza}. I have not addressed the extensive efforts that are ongoing to understand the scale dependence and the matching that is needed to simultaneously understand the behavior at low and high $q_T$ values. Furthermore, links exist with work to understand low-x behavior, use of more involved hadronic observables like di-hadron fragmentation functions or double-parton distributions to understand multi-parton processes, where again we refer to other contributions at this conference.

\acknowledgments
This research is part of the research program of the ``Stichting voor Fundamenteel Onderzoek der Materie (FOM)'', which is financially supported by the ``Nederlandse Organisatie voor Wetenschappelijk Onderzoek (NWO)'' and the EU "Ideas" programme QWORK (contract 320389).


\end{document}